# Assessing the accuracy of the h and g indexes for measuring researchers' productivity[1]

**Giovanni ABRAMO***

*Institute for System Analysis and Computer Science - National Research Council of Italy (IASI-CNR) and*
*Laboratory for Studies of Research and Technology Transfer – University of Rome "Tor Vergata", School of Engineering*
*Via del Politecnico 1,00133 Rome, Italy. E-mail: giovanni.abramo@uniroma2.it*

**Ciriaco Andrea D'ANGELO and Fulvio VIEL**

*Laboratory for Studies of Research and Technology Transfer – University of Rome "Tor Vergata", School of Engineering*
*Via del Politecnico 1,00133 Rome, Italy. E-mail: dangelo@dii.uniroma2.it*

**Abstract**

Bibliometric indicators are increasingly used in support of decisions for recruitment, career advancement, rewarding and selective funding for scientists. Given the importance of the applications, bibliometricians are obligated to carry out empirical testing of the robustness of the indicators, in simulations of real contexts. In this work we compare the results of national-scale research assessments at the individual level, based on three different indexes: the h-index, g-index and "fractional scientific strength", or FSS, an indicator previously proposed by the authors. For each index, we construct and compare rankings lists of all Italian academic researchers working in the hard sciences over the period 2001-2005. The analysis quantifies the shifts in ranks that occur when researchers' productivity rankings by simple indicators such as h- or g-index are compared with that by more accurate FSS.

***Corresponding author**

# 1. Introduction

Bibliometricians are constantly engaged in formulating and improving bibliometric indicators to serve in support of research evaluation. Among the many applications, a very significant one is in evaluation of productivity by individual scientists, for purposes of recruitment, career advancement, selective funding and rewarding. Drawing on citation databases, bibliometrics has times and costs that are very reasonable compared to peer-review, yet permits evaluation of the individual's entire scientific production in a period of time, including its relative impact, proxied by citation counts. However, there are a number of critical methodological issues concerning impact analysis, and in recent years bibliometricians have intensified their efforts to deal with these, as seen in an explosion of theoretical and empirical studies that propose modifications of existing indicators or advance entirely new ones. In 2005, the Argentine American physicist, J.E.Hirsch, achieved an intuitive breakthrough with the proposal of the index that is now named after him (Hirsch, 2005). The "h-index" represents the maximum number $h$ of works by a scientist that have at least $h$ citations each. Hirsch's proposal immediately attracted great international interest because the new indicator represented a single whole number that could synthesize both the quantity and impact of a scientist's portfolio of work. It was precisely the simplicity and ready comprehension of the indicator that determined its success, although this was more with scientists and occasional practitioners than with true bibliometricians. Still, scholars took such interest that citations of Hirsch's original article have exploded to over 1000 and still counting, according to Scopus. Many works took Hirsch's idea, noted the advantages and proposed more or less appropriate applications of the h-index to new analytical contexts: journals, research groups, organizations, countries, etc. (Braun et al., 2006; Van Raan, 2006; Vanclay, 2008; Molinari and Molinari, 2008; Guan and Gao, 2008). Others concentrated on the predictive power of the indicator and attempted to validate its robustness, for application in place of more complex and better known indicators (Hirsch, 2007; Hönekopp and Klebe, 2008; Jensen et al., 2009; Hönekopp and Khan, 2012; Rezek et al., 2011; Carbon, 2011). Many more noted the evident drawbacks and proposed improved variants, leading to a flourishing field of literature on alternative but still "h-like" indicators (Batista et al., 2006; Kosmulski, 2006; Egghe, 2008; Egghe and Rao, 2008; Radicchi et al., 2008; Zhang, 2009; Alonso et al., 2010; Assimakis and Adam, 2010).

The current authors hold that measurement of a scientist's productivity must account for the overall impact of his/her entire production in the period under observation, but the h-index and most of its variants inevitably ignore the impact of works with a number of citations below h and all citations above h of the h-core works, often a very consistent portion, as observed by Ye and Rousseau (2010), and Zhang (2009). The g-index[2] was conceived to take account for the citations above h, but did not solve entirely the h-index limits, because it still neglects all citations outside the g-core works. In measuring impact it is also necessary to consider the specific field (subject category) for each of the scientist's publications and carry out appropriate field-normalization. To this purpose Radicchi et al. (2008) proposed a "generalized h-index", which rescales the number of citations by the average of their distribution in the paper's field. In measuring productivity one should account also for the number of co-authors and their position in the list where it makes a difference. To this end Batista et al. (2006) proposed to divide the h-index of a researcher by the average number of authors in the considered h papers. Last but not least, because of the different intensity of publications across fields, productivity rankings need to be carried out by field (Abramo and D'Angelo, 2007), while it is not unfrequent to resist the temptation to compare the h-indexes of researchers from different fields. Iglesias and Pecharromán (2007) tried to correct this flaw introducing a multiplicative correction to the *h* index which depends basically on the Web of Science (WoS) field the author is in, and to some extent, on the number of papers the researcher has published. Each h-variant indicator tackles one of the many drawbacks of the h-index, while leaving the others unsolved, so none can be considered completely satisfactory. In a previous work we proposed a proxy measure of individual researcher' productivity that meets all the necessary requirements (Abramo and D'Angelo, 2011), which we called the indicator of fractional scientific strength (FSS).

---

[2] The g-index represents the highest number "g" of articles that together received $g^2$ or more citations (Egghe, 2006).

Because the h-index can be easily accessed to in such databases as WoS by Thomson Reuters and Scopus by Elsevier, it is often used to support decisions on recruitment, rewarding and career advancement of scientists. Hirsch himself (2005) recommended the guidelines of h≥12 as a minimum for promotion of a physicist to associate professor and h≥18 for full professor, in leading research universities. Not least as an example, the recent reform of Italian higher education imposes recruitment of associate and full professors by national competitions that are open only to those who exceed threshold levels for certain bibliometric indicators, including the h-index. While the h-index was conceived to characterize the scientific output of a researcher across her/his overall career, it is often applied as a proxy of productivity to compare performance in a given period of time. Given such uses of the indicators, bibliometricians are obligated to empirically test their accuracy, through simulation of real contexts of use. In this work then we compare productivity rankings of Italian academics derived from the FSS, which we use as a benchmark because of its accuracy in measuring productivity, to rankings derived from the h-index and what is probably its best-known variant, the g-index. Our objective is to quantify the levels of accuracy for the h and g indexes when applied to evaluate productivity at the level of the individual scientist and obtain useful information on whether and to what extent the g-index represents an improvement of the h-index in measuring productivity.

Analysis of the literature reveals that one of the characteristics of studies on the theme, regardless of the ultimate objective, is that they involve quite narrow fields of observation. Theoretical studies are largely based on fictitious examples (Bornmann and Daniel, 2007; Marchant, 2009; Hirsch, 2010; Bouyssou and Marchant, 2011; Waltman and Van Eck, 2012), while empirical analyses generally refer to limited sets of fields or institutions (Bormann et al., 2008; Costas and Bordons, 2008; Van Raan et al., 2010).

The main difficulty in conducting large scale measurement of the h-index, as for any other bibliometric indicator of productivity, involves the occurrence of significant problems of homonyms in large populations of scientists. Eliminating ambiguities as to the precise identity of the author within acceptable margins of error is a daunting task. For this reason Bornmann and Daniel (2007) recommend "calculating the *h-index* on the basis of a complete list of publications that is authorized by the scientist himself or herself". However application of this recommendation on large scale, within a reasonable budget, is obviously unfeasible. To counter past shortcomings, Abramo et al. (2010) developed an unambiguous data set, within acceptable margins of error, to carry out a large scale measurement exercise of the h and g indexes. They provided descriptive statistics concerning over 20,000 Italian academic scientists working in 165 subject fields, offering robust benchmarks for comparing individual productivity in the same subject field. Based on this experience, the authors now propose to compare the results from large-scale evaluation exercises at the level of individual scientists using alternative indicators of productivity. We compare the national rankings lists derived from values of h-index, g-index and FSS, for the scientific production over the 2001-2005 period by all Italian academic researchers in the hard sciences.

A review of seemingly related literature to compare findings is not simple, for two reasons: i) works that at first appear analogous actually have much different aims and objectives; ii) results do not always converge, even when originating from authoritative publications in the field. Jensen et al. (2009), for example, affirm to have shown that, overall, h-index is the best bibliometric indicator to account for the promotions of about 600 researchers at France's CNRS. However, they compare h-index to indicators that are equally simple and imperfect, i.e. number of publications; number of citations; mean citations per paper; ratio of h-index to number of papers. Ball (2007) reached similarly positive conclusions, affirming that the h-index does seem able to identify good scientists. However there are more than a few scholars with conflicting opinions. A prime example is the contribution by Marchant (2009), who argues that the adequacy of an indicator must be evaluated on the basis of its context of use. Yet the h-index certainly violates certain axiomatic properties (in particular the principles of independence and weak independence), which bibliometric indicators should always possess: in consequence, there are many contexts where rankings based on h-index cannot be reasonable. A second contribution that again exemplifies this position, but on an empirical basis, is by Bornmann et al. (2008): these authors assume peer reviews as benchmark for selection decisions on research fellowships in biomedicine, and find that indicators other than h-index are even better suited for the evaluation purposes.

Section 2 presents the methodology for the proposed analysis and the characteristics of the dataset. Section 3 provides the results from comparisons of the rankings constructed using the different indicators considered. Section 4 provides an in-depth analysis of a specific subset of researchers: those who place at the top of the rankings, and who are thus of greater interest for recruitment, career advancement and selective funding. The final section summarizes the results of the work, compares them to previous assertions in the literature, and discusses the implications.

## 2. Methodology and dataset

The bibliometric dataset used in the analysis is extracted from the Italian Observatory of Public Research, a database developed and maintained by the authors and derived under license from the WoS. Beginning from the raw data of the WoS, and applying a complex algorithm for reconciliation of the author's affiliation and disambiguation of the true identity of the authors, each publication is attributed to the university scientist or scientists that produced it (D'Angelo et al., 2011).

The proposed analysis is based on publications (articles, reviews and conference proceedings) authored by Italian academic scientists in the period 2001-2005. The period is sufficiently long to avoid randomness in the scientific production and guarantee robustness of the measures. Citations are observed as of 30/06/2009, providing a sufficient citation window to guarantee a reliable impact assessment. We take advantage of a unique feature of the Italian university system, in which all research personnel are classified in one and only one scientific field. In the hard sciences there are 205 such fields (named scientific disciplinary sectors, SDSs), grouped into nine disciplines (named university disciplinary areas, UDAs[3]). To assure full representativeness of publications as proxy of the research output, the field of observation is limited to those SDSs (184 in all, accessible at http://www.disp.uniroma2.it/laboratoriortt/TESTI/Indicators/ssd2.html) where at least 50% of researchers produced at least one publication in the observed period.

The identification of the research staff and their SDS classifications, for each university, is accomplished by referring to a database on all Italian personnel maintained by the Ministry of Universities and Research. In the five years under consideration, there were 35,002 scientists (assistant, associate and full professors) on staff in the 184 SDSs considered. To assure greater reliability, the analysis excludes all professors who entered or left the university system during the period of observation. Thus the final dataset is reduced to 28,219 scientists: their distribution by UDA is shown in Table 1. Over the five years considered, they authored a total of approximately 136,000 publications, receiving over 1.6 million citations by 30/06/2009.

[Insert Table 1 here]

For each scientist, we measure his/her research productivity by the three indicators h-index, g-index and FSS. Based on the three indicators, each scientist is compared to all other Italian colleagues in the same SDS and rankings are provided. We exclude 4,548 professors with no publications and 1,225 with no citations over the period, since their rank is not dependent on the selected indicator.

---

[3] Mathematics and computer sciences; physics; chemistry; earth sciences; biology; medicine; agricultural and veterinary sciences; civil engineering; industrial and information engineering.

Research activity is a production process in which the inputs consist of human, tangible (scientific instruments, materials, etc.), and intangible (accumulated knowledge, social networks, etc.) resources; and where output, i.e. the new knowledge, has a complex character of both tangible nature (publications, patents, conference presentations, databases, protocols, etc.), and intangible nature (tacit knowledge, consulting activity, etc.). The new-knowledge production function has therefore a multi-input and multi-output character. The principal efficiency indicator of any production system is productivity. When measuring productivity at the individual level, if there are differences in the production factors (capital, scientific instruments, materials, etc.) available to each scientist then one should normalize by them. Unfortunately, in Italy relevant data are not available at individual level. The first assumption then, is that resources available to researchers within the same field of observation are the same. The second assumption is that the hours devoted to research are more or less the same for all researchers. These assumptions are fairly well satisfied in the Italian higher education system, which is mostly public and not competitive. Up to 2009, the core funding by government was input oriented, meaning that it was distributed to universities in a manner intended to satisfy the needs for resources of each and all, in function of their size and activities. Furthermore, the time to devote to education is established by law.

To assess productivity of individual researchers by FSS, we consider the outcome, or impact of their research activities. As proxy of outcome we adopt the number of citations for the researcher's publications. Because the intensity of publications varies by field, we compare researchers within the same field, meaning the same SDS. Another issue is that it is very possible that researchers belonging to a particular scientific field will also publish outside that field. Because citation behavior varies by field, we standardize the citations for each publication with respect to the median of the distribution of citations for all the Italian cited-only publications[4] of the same year and the same WoS subject category. Furthermore, research projects frequently involve a team of researchers, which shows in co-authorship of publications. In this case we account for both the fractional contributions of scientists to outputs, as the reciprocal of number of co-authors, and their position in the list[5]. The productivity of a single researcher by FSS, is given by:

$$FSS = \sum_{i=1}^{n} \frac{c_i}{m_i} * \frac{1}{s_i}$$

Where:

$c_i$ = citations received by publication *i*;

$m_i$ = median of the distribution of citations received for all Italian cited-only publications of the same year and subject category of publication *i*;

$s_i$ = co-authors of publication *i*

n = number of publications of the researcher in the period of observation.

In the life sciences, widespread practice is for the authors to indicate the various contributions to the published research by the positioning of the names in the authors list. For the life sciences then, when the number of co-authors is above two, different weights are given to each co-author according to his/her position in the list and the character of the co-authorship (intra-mural or extra-mural). If first and last authors belong to the same university, 40% of citations are attributed to each of them; the remaining 20% are divided among all other authors. If the first two and last two authors belong to different universities, 30% of citations are attributed to first and last authors; 15% of citations are attributed to second and last author but one; the remaining 10% are divided among all others[6].

---

[4] We refer to Italian publications because the world median of citations is not made available, unless one buys all world WoS data. We take into account cited-only publications, otherwise the median would be nihl in a number of WoS subject categories.

[5] For life sciences, different weights are given to each co-author according to his/her position in the list and the character of the co-authorship (intra-mural or extra-mural). If first and last authors belong to the same university, 40% of citations are attributed to each of them; the remaining 20% are divided among all other authors. If the first two and last two authors belong to different universities, 30% of citations are attributed to first and last authors; 15% of citations are attributed to second and last author but one; the remaining 10% are divided among all others.

FSS is similar to the Leiden CWTS new crown indicator, the "mean normalized citation score" (Waltman et al., 2011), and the Lundberg's (2007) "item-oriented field-normalized citation score average". The last two refer to the evalutation of average impact of a set of publications, while FSS to the evaluation of productivity.

Because both the h-index and the g-index ignore, although to a different extent, part of the overall impact of a researcher's output, and neither normalize citations by field, or take into account the number of co-authors and their position in the list, such indexes are less accurate than FSS in measuring productivity, the main indicator of efficiency in any production process.

## 3. Results

We carry out the comparison of the rankings lists from the three indicators by a series of steps. First we present the case of a single SDS, then extend the analysis to all SDSs of a UDA and finally to all hard science UDAs. Table 2 shows the example of the MED/31-Otorinolaringology SDS, of the medicine UDA. This SDS had 132 professors in stable role over the five-year period. Thirty-one of these did not achieve any publications and another seven, while having published, were never cited. For the remaining 94, columns 2, 3 and 4 show the absolute value of the three indexes of productivity, and the next three columns show their corresponding quartile ranks. ID numbers are assigned according to FSS rank. The last two columns show the value of the quartile shift between the FSS ranking and the rankings from h and g indexes. For the first three professors of the list there is no variation in productivity: their scientific production over the five years places them in the first quartile for productivity no matter what indicator is considered. However, running down the lists, we begin to observe movements: for ID_25, which is the researcher with the highest FSS of the second quartile, we note that both the h and g-index values place the individual in the higher first productivity quartile, while it does not occur for researchers further down the list. Overall, over a third of the scientists (exactly 34[7] out of 94) show a different quartile of productivity under evaluation with the h-index and with FSS, and an analogous number (35 of 94) show variation in quartile under the g-index and FSS. For two scientists, the shift in rankings between h-index and FSS is a full two quartiles, and the same double shift of quartile occurs, again for two scientists, between the rankings by g-index and FSS.

---

[6] The weighting values were assigned following advice from Italian professors in the life sciences. The values could be changed to suit different practices in other national contexts.

[Insert Table 2 here]

We repeat the same type of measurement for all the SDSs of each UDA. As an example, we present the analysis for the chemistry UDA, which has 12 SDSs. The CHIM/05 SDS was excluded because, in the five years under examination, there were only three research staff in the entire nation. For each of the remaining SDSs of the UDA we measure the correlation between the productivity ranking list for FSS and the lists from h and g-index. In Table 3, we observe that there is a very high correlation between the rankings. The first comparison, FSS versus h-index, returns correlation values that are constantly greater than 0.83, and higher than 0.89 in seven of the 11 SDSs analyzed. The comparison between g-index and FSS presents a highly similar situation.

---

[7] Not to be confused with the "total" indicated in Table 2, which representes the sum of the quartile shifts: two scientists register double shifts in quartile.

[Insert Table 3 here]

However, such high values of correlation can still hide very substantial variations in ranking at the level of individual researchers. Table 4 presents the example of the descriptive statistics for variations in quartile for the researchers of each of the chemistry SDSs. We note that on average, roughly a third of each SDS's research staff show different quartiles of productivity when they are evaluated with indicators other than FSS (columns 3 and 4). In CHIM/08 (Pharmaceutical chemistry) the percentage of staff registering quartile differences between the h-index and FSS rankings is actually 45%, and differences remain high (43% of staff) in comparing the g-index and FSS rankings. The average shift in quartile for the 369 researchers in this SDS is the highest for the UDA, at 0.5 for the h-index comparison and 0.46 for g-index. There are even cases of researchers with shifts of three quartiles, meaning that they rank first (or last) when evaluated by FSS and then last (or first) for one of the other two indicators. The CHIM/09 SDS (Applied Technological Pharmaceutics) also shows important differences in rankings, with 40-41% of researchers showing a different productivity quartile under different indicators, with an average quartile difference of 0.47 for the h-index to FSS comparison and 0.43 for g-index to FSS. In all the other SDSs, the comparison between rankings shows lesser values of differences, but still very meaningful. The largest SDS, CHIM/06 (Organic chemistry), has the smallest shifts. Still, there are changes in ranking for roughly a quarter of the total research staff (557 scientists), with the shifts in quartile between the h and g indexes and the FSS rankings averaging 0.26 and 0.27 respectively.

Now we extend the analysis to all the UDAs: Figure 1 presents the distribution of Spearman correlation index for rankings based on h or g-index, on the one hand, and FSS on the other hand, for all 182 SDSs considered[8]. The substantial superimposition of the two curves is clear, even though the curve for the FSS to g-index comparison is almost always above the FSS versus h-index curve. For this latter comparison, there are 38 (or 21% of total) SDSs that show correlation greater than 0.9 and a full 150 (82% of total) with correlation greater than 0.8. There are eight SDSs that show below 0.6 and only two below 0.4 (AGR/10- Rural Construction and Environmental Land Management; ING-IND/02- Naval and Marine construction and installation). For the comparison between the g-index and FSS rankings, we note that there are 49 SDSs with correlation greater than 0.9, while for all the other thresholds the numbers are similar to those for the h-index to FSS comparison.

[Insert Table 4 here]

---

[8] In addition to CHIM/05, MED/48 (Neuropsychiatric and Rehabilitation Nursing) is also excluded from this analysis, again for reasons of the limited number of observations.

[Insert Figure 1 here]

*Figure 1: Spearman correlation for rankings based on g / h-index and FSS: distribution by SDSs*

The analysis of quartile variations shows important differences between the two comparisons (Figure 2). For the majority of SDSs analyzed (88 of 182), the shifts in quartile for h-index and FSS concern between 40% and 60% of the researchers. However in comparing quartiles for g-index and FSS, the modal class is the 20% to 40% group: for 103 of the 182 SDSs the percentage of researchers involved in quartile shifts falls between these limits. Evidently the differences in quartile rankings by g-index and FSS are less numerous than those by h-index and FSS. Table 5 shows the variability of the situations encountered in the SDSs of the individual UDAs: for each UDA, the table presents the descriptive statistics concerning shifts in quartile for the two SDSs with the maximum and minimum percentage of researchers registering variations in the rankings from h-index and FSS. Where notable differences emerge between SDSs within the same UDA, for example in Industrial and information engineering, they can be ascribed to the concurrence of three factors: i) different citation behaviors across SDSs; ii) different intensity of publications; and iii) different collaboration rates. In fact, differently from FSS, both h and g indexes neglect to normalize by the above factors. Those individual scientists registering 3 quartile variations are those who tend to publish with low number of co-authors, in subject categories with low citation rates and are consequently penalized by performance indicators which do not account for such differences.

[Insert Figure 2 here]

*Figure 2: Number of SDSs per percentage interval of researchers registering quartile variations*

[Insert Table 5 here]

Table 6 presents a synthesis of statistics comparing the rankings constructed with the three bibliometric indicators considered. The statistics are obtained by UDA, aggregating the data on the researchers of their constituent SDSs. The Spearman correlation takes the lowest value in the physics UDA, both for the h-index/FSS comparison (0.68) and for g-index/FSS (0.67). The highest value (0.90), constant for the two comparisons, is registered for the chemistry UDA. However there is a greater correlation between the FSS and g-index rankings in five out of the nine UDAs: civil engineering; industrial and information engineering; agricultural and veterinary science; earth sciences; mathematics and computer science. In the remaining four UDAs (biology, chemistry, physics, medicine) the levels of accuracy for the h and g indexes are equivalent, both in terms of correlation between the rankings (columns 2 and 3), and in average quartile variation (columns 4 and 5).

[Insert Table 6 here]

In comparing the rankings for FSS and h-index, the overall percentage of researchers registering a quartile variation is 41% (last line, second column, Table 7), with a peak in civil engineering (48.6%) and a minimum in chemistry (33.8%). For the g-index/FSS comparison, the distribution of values decreases significantly: the overall average of researchers registering a quartile variation is 37.3% (last line, column 3, Table 7), with a minimum (31.7%) in chemistry and a maximum (47.1%) in physics.

[Insert Table 7 here]

Columns 4 and 5 of Table 7 also reveal the number of cases characterized by notable shifts in rank, of two quartiles or more. Such shifts concern an average of 4.4% of researchers in the comparison between rank by h-index and by FSS and 3.8% in the g-index/FSS comparison. The most notable percentage occurs in physics: if evaluated by g-index, 10.2% of researchers in this UDA would have a position much different from that under evaluation by FSS. The number of such cases is much more limited in all the other UDAs, and especially in chemistry (1.8%).

In summary, the comparisons reveal a strong correlation in the rankings obtained from the bibliometric indicators considered. The g-index seems more correlated to FSS than the h-index does, in at least five UDAs. In the other four, physics included, the differences in rankings for the h and g indexes relative to the FSS rankings are almost identical. However, at the level of individual researchers, the percentage of those affected by shifts in quartile, when evaluated by h and g indexes, is certainly significant, as is the average value of such shifts. The problem is particularly notable in physics, meaning that this is a UDA where the choice of the bibliometric indicator seems particularly critical.

Now we ask whether the extent of shifts that we have seen remain similar when, instead of referring to all researchers in an SDS, we examine only the top scientists, or those that place in the first quartile of national rankings for their SDS: a highly interesting subgroup for issues of recruitment, career advancement and selective funding. The next section provides an in-depth analysis.

**4. Analysis of top scientists**

We begin by using the three chosen indicators to identify the "top scientists" in each SDS, meaning those that place in the first quartile for productivity. We then compare the three subsets of top scientists thus identified, first measuring the extent of their intersection. Figure 3 summarizes this analysis for the example of the MED/31 SDS.

[Insert Figure 3 here]

*Figure 3: Intersections between the subsets of top scientists as identified on the basis of each indicator (h-index, g-index, FSS) for SDS MED/31*

Among the top scientists ranked for h-index, 92% coincide with those scored for g-index, and 81% with those for FSS (left chart). The central chart indicates that all top scientists as ranked under g-index also achieve the first productivity quartile for h-index, while only 79% result as "top" for FSS. Finally, in 12% of cases, the top scientists by FSS would not achieve top if evaluated on the basis of h-index and in 21% of cases they would not achieve top if evaluated by g-index.

As in the preceding section, we extend the analysis to all the SDSs of each UDA. We consider the researchers that result as top for FSS in each SDS, and we verify how many, on the basis of h and g indexes, would lose this attribute. The analysis to all 182 SDSs under examination permits an appreciation of the differences between the overall disciplinary areas: Table 8 presents the data obtained by aggregation of each SDS into the composite UDAs. The last line of the table shows that, on average, 17% of the top scientists as ranked by FSS do not achieve this level under h-index. Physics is definitely the most problematic UDA: 36% of top scientists by FSS are not at the top of rankings derived from h-index. In the other UDAs this percentage is always less than 20%, but with the sole exception of civil engineering it is still greater than 10%. In four UDAs (industrial and information engineering, biology, physics, medicine) we observe cases of researchers with jumps of three quartiles: these are researchers classified as top scientist for FSS but in the last quartile for h-index, or vice versa.

[Insert Table 8 here]

The comparison between top scientists for FSS and those for g-index shows shifts that are still greater than those for h-index. In general, 21% of top researchers for FSS do not reach the first productivity quartile for g-index (last line, fourth column of Table 9). Physics is again the most problematic UDA, with 42% of top scientists for FSS not achieving top for g-index. Observing that, differently form the h- and g-indexes, FSS takes into account: i) the impact of works with a number of citations below h and all citations above h of the h-core; and ii) the number of co-authors, there are two factors then which may concur to such notable differences. First, Physics is a discipline with very high intensity of publications and citations (Abramo and D'Angelo, 2009). Second, the number of co-auhtors is generally very high, in many SDSs within Physics and especially in Nuclear Physics. There are five cases of variations of three quartiles. In addition to what we have seen in Section 3 there is thus a significant new feature: again comparing to FSS-based evaluation as benchmark, it is evident that for top performers, evaluation conducted by g-index differs more than evaluation conducted by h-index.

## 5. Discussion and conclusions

One of the pressing issues currently engaging bibliometricians concerns the formulation of appropriate indicators in support of decisions on recruitment, career advancement, selective funding and rewarding of individual scientists. Those who are directly concerned, namely the researchers, demand that any systems of evaluation for their productivity, regardless of simplicity, be transparent, exhaustive and trustworthy. On the other hand, the success and widespread use of indicators such as the h-index and its well-known variant, the g-index, highlights how the need for administrative efficiency often push practitioners to adopt simple evaluation systems and indicators.

In this work, we proposed assessment of the accuracy of the h and g indexes for measuring researchers' productivity, considering a third index as benchmark: *fractional scientific strength*, an indicator that measures the impact of a researcher's entire scientific production in a period of time, not just that of the most cited publications; normalizes citations by field; and accounts for the number of authors who contributed to the publication.

The results from the current work diverge from those by Jensen et al. (2009) and Ball (2007), while seem aligned with the position exemplified by the works of Marchant (2009) and Bornmann et al. (2008). One of the novel elements in our study is certainly the scale of the empirical analysis undertaken, at the level of an entire national university system. The analysis reveals a high correlation between the rankings obtained from the three bibliometric indicators considered. In comparison between FSS and the h-index in 38 fields (SDSs) of a total 182, the correlation between the two rankings is greater than 0.9, and there are a full 150 SDSs with correlation greater than 0.8.

However these high levels of correlation conceal very substantial variations in rankings at the level of individual researchers.

In the comparison between h-index and FSS, the overall share of researchers registering a quartile variation is 41%, with a maximum peak in civil engineering (48.6%) and a minimum in chemistry (33.8%). Jumps of a quartile between h-index and FSS affect shares of between 40% and 60% of the total researchers, for a full 88 SDSs out of 182.

Further, cases of remarkable shifts in rank, of two or more quartiles, are not at all isolated: on average, such shifts concern 4.4% of researchers in the comparisons of rankings from h-index and FSS, and 3.8% of comparisons between g-index and FSS. The most notable percentage occurs in physics (10.2%), and the lowest is in chemistry (1.8%).

An analysis focused on the leaders of the rankings lists is definitely of interest, for issues of recruitment, career advancement and selective funding. It reveals that, on average, 17% of the top scientists for FSS do not achieve the first national quartile for h-index. It is again physics (J.E. Hirsch's research area) that appears as most problematic: over a third of scientists that belong to the first quartile for FSS productivity do not reach "top" in rankings for h-index. In the other UDAs this percentage is always below 20%, but with the exception of civil engineering it is still always above 10%.

Taking FSS as benchmark, the percentages of researchers affected by jumps in quartile in the comparison to h-index and g-index reach percentages that cannot be ignored, just as the average value of these shifts also cannot be ignored. The problem is particularly relevant in physics, where the choice of bibliometric indicator is particularly critical, whether referred to the entire population or focused only on top scientists.

As it has already been observed empirically, when the difference between h-indices is large enough, the h-indices usually reflect their performance difference. However, if the difference is small or zero, the h-indices would fail to distinguish performance difference (Kuan et al., 2011) and therefore big shifts are certainly possible. Although correlations of rankings by the above indicators are very high, shifts in ranks of individual researchers, should put the operator on guard over the temptation to adopt in any circumstances simple indicators for evaluation of bibliometric productivity of individual scientists. While they are easy to understand, quantify and communicate, such indicators conceal a level of inaccuracy in measuring research productivity that is generally unacceptable for most of the intended uses and objectives. Our recommendation then is to avoid the use of the h-index and its variances in comparative assessments of research productivity.

|  | | | h index quartiles | | | | |
| --- | --- | --- | --- | --- | --- | --- | --- |
| UDA | N. of SDS | N. of scientists | 1° | Median | 3° | Max | Average | Variance |
| Mathematics and computer sciences | 9 | 1,732 | 1 | 2 | 3 | 13 | 2.31 | 2.23 |
| Physics | 7 | 1,846 | 2 | 4 | 7 | 25 | 5.04 | 12.50 |
| Chemistry | 11 | 2,597 | 4 | 6 | 8 | 36 | 6.24 | 13.74 |
| Earth sciences | 12 | 794 | 1 | 2 | 4 | 11 | 2.90 | 4.10 |
| Biology | 19 | 3,621 | 3 | 4 | 7 | 33 | 5.02 | 12.64 |
| Medicine | 41 | 6,277 | 2 | 4 | 7 | 33 | 4.94 | 15.56 |
| Agricultural and veterinary sciences | 25 | 1,514 | 1 | 2 | 4 | 18 | 3.03 | 4.88 |
| Civil engineering and architecture | 5 | 484 | 1 | 2 | 3 | 12 | 2.41 | 2.89 |
| Industrial and information engineering | 36 | 2,573 | 1 | 2 | 4 | 19 | 2.89 | 4.63 |

*Table 1: h index quartiles for Italian university scientists grouped by UDA*

|  | g index quartiles | | | | | |
| --- | --- | --- | --- | --- | --- | --- |
| UDA | 1° | Median | 3° | Max | Average | Variance |
| Mathematics and computer sciences | 1 | 3 | 4 | 47 | 3.38 | 9.56 |
| Physics | 3 | 6 | 11 | 43 | 7.75 | 35.45 |
| Chemistry | 5 | 8 | 12 | 55 | 9.18 | 34.39 |
| Earth sciences | 2 | 3 | 6 | 18 | 4.12 | 10.63 |
| Biology | 3 | 6 | 10 | 58 | 7.37 | 32.40 |
| Medicine | 3 | 6 | 11 | 58 | 7.62 | 46.63 |
| Agricultural and veterinary sciences | 2 | 3 | 6 | 35 | 4.52 | 13.44 |
| Civil engineering and architecture | 1 | 3 | 5 | 20 | 3.45 | 7.50 |
| Industrial and information engineering | 2 | 3 | 6 | 32 | 4.24 | 12.83 |

*Table 2: g index quartiles for Italian university scientists grouped by UDA*

|  |  | h index quartiles | | | | | |
| --- | --- | --- | --- | --- | --- | --- | --- |
| SDS | N. of scientists | 1° | Median | 3° | Max | Average | Variance |
| FIS/01 | 745 | 2 | 4 | 6 | 22 | 4.50 | 10.41 |
| FIS/02 | 264 | 2 | 5 | 7 | 17 | 5.14 | 11.06 |
| FIS/03 | 331 | 4 | 6 | 8 | 25 | 6.29 | 14.00 |
| FIS/04 | 133 | 2 | 4 | 6 | 11 | 4.32 | 7.57 |
| FIS/05 | 134 | 3 | 5 | 10 | 23 | 6.91 | 28.59 |
| FIS/06 | 42 | 2 | 3 | 4 | 10 | 3.21 | 4.12 |
| FIS/07 | 197 | 2 | 4 | 6 | 13 | 4.45 | 6.83 |

*Table 3: h index quartiles for Italian university scientists in the Physics UDA*

|  |  | g index quartiles | | | | | |
| --- | --- | --- | --- | --- | --- | --- | --- |
| SSD | N. of scientists | 1° | Median | 3° | Max | Average | Variance |
| FIS/01 | 745 | 3 | 6 | 10 | 37 | 6.99 | 28.47 |
| FIS/02 | 264 | 3 | 7 | 11 | 30 | 7.78 | 33.74 |
| FIS/03 | 331 | 5 | 9 | 12 | 43 | 9.79 | 44.20 |
| FIS/04 | 133 | 3 | 6 | 11 | 22 | 6.83 | 25.52 |
| FIS/05 | 134 | 4 | 8 | 16 | 36 | 10.52 | 73.18 |
| FIS/06 | 42 | 2 | 4 | 6 | 17 | 4.64 | 14.09 |
| FIS/07 | 197 | 3 | 6 | 9 | 20 | 6.55 | 17.66 |

*Table 4: g index quartiles for Italian university scientists in the Physics UDA*

|  | Median | | Max | |
|---|---|---|---|---|
| UDA | Min | Max | Min | Max |
| Mathematics and computer sciences | 1 | 2 | 6 | 13 |
| Physics | 3 | 6 | 10 | 25 |
| Chemistry | 3 | 6 | 7 | 36 |
| Earth sciences | 1 | 4 | 6 | 11 |
| Biology | 1 | 6 | 7 | 33 |
| Medicine | 2 | 7 | 6 | 33 |
| Agricultural and veterinary sciences | 1 | 5 | 6 | 18 |
| Civil engineering and architecture | 2 | 3 | 6 | 12 |
| Industrial and information engineering | 1 | 5 | 4 | 19 |

*Table 5: Ranges of medians and maximums for the distribution of h indexes among the SDSs of each UDA*

| UDA | Total N. of SDSs | N. of these with first quartile = 1 | N. of these with median <= 2 |
|---|---|---|---|
| Mathematics and computer sciences | 9 | 9 | 9 |
| Physics | 7 | 0 | 0 |
| Chemistry | 11 | 1 | 0 |
| Earth sciences | 12 | 7 | 6 |
| Biology | 19 | 2 | 1 |
| Medicine | 41 | 13 | 10 |
| Agricultural and veterinary sciences | 25 | 14 | 15 |
| Civil engineering and architecture | 5 | 4 | 4 |
| Industrial and information engineering | 36 | 24 | 22 |

*Table 6: Number of SDSs where the first quartile of h index equals 1 and the median is less than or equal to 2, for each UDA*

|  | Median | | Max | |
|---|---|---|---|---|
| UDA | Min | Max | Min | Max |
| Mathematics and computer sciences | 2 | 3 | 10 | 47 |
| Physics | 4 | 9 | 17 | 43 |
| Chemistry | 5 | 9 | 10 | 55 |
| Earth sciences | 2 | 5 | 8 | 18 |
| Biology | 2 | 8 | 10 | 58 |
| Medicine | 2 | 11 | 9 | 58 |
| Agricultural and veterinary sciences | 2 | 7 | 9 | 35 |
| Civil engineering and architecture | 2 | 4 | 11 | 20 |
| Industrial and information engineering | 1 | 7 | 7 | 32 |

*Table 7: Range of medians and maximums for the distribution of g indexes among the SDSs of each UDA*

| UDA | Total N. of SDSs | N. of these with first quartile = 1 | N. of these with median <= 2 |
|---|---|---|---|
| Mathematics and computer sciences | 9 | 5 | 4 |
| Physics | 7 | 0 | 0 |
| Chemistry | 11 | 1 | 0 |
| Earth sciences | 12 | 5 | 3 |
| Biology | 19 | 2 | 1 |
| Medicine | 41 | 6 | 1 |
| Agricultural and veterinary sciences | 25 | 9 | 5 |
| Civil engineering and architecture | 5 | 2 | 2 |
| Industrial and information engineering | 36 | 13 | 12 |

*Table 8: Number of SDSs where the first quartile of g index equals 1 and the median is less than or equal to 2, for each UDA*